# Understanding of phase noise squeezing under fractional synchronization of non-linear spin transfer vortex oscillator

R. Lebrun[1], A. Jenkins[1], A. Dussaux[1*], N. Locatelli[1+], S. Tsunegi[1,2], E. Grimaldi[1], H. Kubota[2], P. Bortolotti[1], K. Yakushiji[2], J. Grollier[1], A. Fukushima[2], S. Yuasa[2] and V. Cros[1]

[1]Unité Mixte de Physique CNRS/Thales and Université de Paris Sud Palaiseau, France.
[2] Spintronics Research Center, National Institute of Advanced Industrial Science and Technology (AIST), Tsukuba, Japan

We investigate experimentally the synchronization of a vortex based spin transfer oscillator to an external rf current whose frequency is at multiple integers, as well as half integer, of the oscillator frequency. Through a theoretical study of the locking process, we highlight both the crucial role of the symmetries of the spin torques acting on the magnetic vortex and the nonlinear properties of the oscillator on the phase locking process. Through the achievement of a perfect injection locking state, we report a record phase noise reduction down to -90dBc/Hz at 1 kHz offset frequency. The phase noise of these nanoscale oscillators is demonstrating as being low and controllable which is of significant importance for real applications using spin transfer devices.

In the last decade, large expectations have been anticipated on how the rich spin transfer physics will give birth to a new generation of multi-functional spintronic devices [1]. The tunable response of spin torque devices has been predicted to play a crucial role in several domains such as radio frequency [2] or magnonic [3] nanoscale and low energy cost devices for ICTs as well as neuro-inspired memory devices [4]. For all of these potential applications, and notably for the corresponding microwave applications, it is essential to identify the mechanisms leading to a fine control of the phase of these spin torque devices. Indeed, it has been often emphasized that their nonlinear behavior gives a unique opportunity to tune their radiofrequency properties [5–7] but at the cost of large phase noise, not compatible with targeted applications [1,2]. In order to tackle these issues, a solution is to rely either on their synchronization to a reference external signal [8–11] or to achieve mutual synchronization [12,13] in arrays of spin torque nano-oscillators (STNOs). However, in all the reported studies made on the locking regime of STNOs, the phase noise, often measured through the estimation of the spectral linewidth measured with a spectrum analyzer, remains large, typically in the kHz range. This feature reveals that phase slips associated with the large thermal energy lead to a loss of synchronization [9,10,14] and have a detrimental and non-controllable impact on the expected behavior of STNOs.

In this letter, we investigate the mechanism leading to a perfect phase locking of a double vortex based STNO to an external rf current with a $F_s$ frequency at $f_0/2$, $f_0$ and $2f_0$, where $f_0$ is the frequency of our STNO. Indeed, thanks to their large intrinsic coherence compared to other types of STNOs [7,15,16], we succeed to elucidate the strong correlation between the oscillator parameters and the locking process through a thorough experimental study combining time domain measurements and analytical developments. This allows understanding of the locking range characteristics [8,17,18] as well as the high phase coherence in the locked regime [19–21]. Our results demonstrate the specific spin transfer locking process of our vortex based STNO, and its potential outcomes for synchronizing multiple oscillators in series, which is an important breakthrough toward rf devices or associative memory applications [1].

All the presented high frequency transport measurements have been performed at room temperature on circular hybrid magnetic tunnel junctions (MTJ) (with 300 or 400 nm diameters Ø) patterned from magnetic multilayered stacks which consist of CoFe / Ru / CoFeB / MgO / NiFe (6nm) / Cu / NiFe (20nm). Details about the sample fabrication can be found elsewhere [7]. The magnetization configuration of each NiFe layers is at remanence a magnetic vortex state. We previously demonstrated that the excitation by mutual spin transfer torque of coupled vortex modes inside the NiFe/Cu/NiFe trilayers is detected at zero applied magnetic field [7,15,16]. Interestingly, this excited coupled mode displays very narrow linewidth (~100 kHz) [7,15,16]. In the samples studied here, the bottom part of the devices i.e. the CoFe/Ru/CoFeB/MgO stack permits the probing of the dynamics of the excited vortex in the 6 nm NiFe layer and thus the output emitted power of the device is directly proportional to the magneto-resistive ratio of our junctions (~ 80%).

Here we study the injection locking features of strongly coherent vortex oscillations in the self-sustained regime with an external rf current. In Fig. 1, we observe a large locking range when the frequency of the current source $F_s$ is chosen to be close to the oscillator frequency $f_0$ (Fig. 1.b). Beyond this expected behavior, we observe that a frequency locking can be also achieved when the condition $f_0 = qF_s$ is fulfilled with $q$ an integer (1 in Fig. 1.b, 2 in Fig. 1.c; see also 3 in supplementary information [22]). Such synchronization at higher harmonics ($q$ integer) [23] is usually associated to the presence of harmonics in the autonomous regime [24] and thus is less efficient as $q$ increases (as the amplitude of harmonics decreases with $q$).

Importantly, as displayed in Fig. 1a, we prove, for the first time in the case of STNO, the synchronization on the external rf current even for a fraction of the frequency ($f_0/q$), here at $F_s = f_0/2$. It should be noted that sub-harmonic phase locking has been predicted in micromagnetic calculations [18] but never observed experimentally. Indeed, sub-harmonic synchronization ($1/q$) generally relies only on the nonlinear current-voltage relationship that, in turn, generates a small voltage signal around $f_0$ [25]. Thus, as shown in Fig. 1, we find that the locking range is much smaller at $f_0/2$ (around 1% of $f_0$) compared to the ones at $f_0$ and $2f_0$ (respectively 10% and 5%). The ability of our double vortex based STNO to synchronize strongly to an external signal is further illustrated by a significant improvement of the spectral coherence when phase locking is achieved. Indeed we observe, both for sub and higher harmonics synchronization, a perfect phase locking state with a minimum

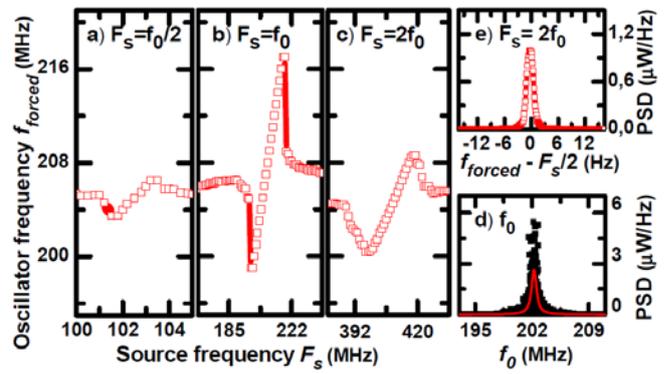

Fig. 1 : Frequency of STNO vs external frequency at $f/2$ (a), $f$ (b), $2f$ (c) for $I_{rf}$ = 0.8 mA (at zero applied field and $I_{dc}$ = + 16 mA and Ø = 400 nm). (d) (e) Emitted spectrum in the autonomous regime and in the synchronized state for $F_s = 2f_0$

linewidth of 1 Hz as shown in Fig. 1.e (limited by the resolution band width of the spectrum analyzer) i.e. about $10^5$ lower than the autonomous regime (700 kHz, Fig. 1.d). Moreover this strong linewidth reduction is combined with a total output emitted power increased by 30% which reaches up to 1.1 μW, which may be associated with an increase of the radius of gyration [22,26]. We believe that such a level of performance for spin transfer oscillators represents a breakthrough towards the actual development of new generations of injection-locked frequency divider/multipliers.

To gain a deeper understanding of the main characteristics of phase noise when the oscillator is synchronized, we perform time domain measurements by recording 5 ms long output voltage time traces with a single-shot oscilloscope (for details see [26]). As already reported for other STNOs [26,27], we find that the power spectral density (PSD) of the phase noise in the autonomous regime displays a $1/f^2$ dependence from the carrier frequency that is associated to a white frequency noise (see black curves in Fig. 2). This phase noise behavior is seemingly modified when an external current $I_{rf}$ is applied at $f_0$, at $f_0/2$ or $2f_0$. When $F_s = f_0$, a significant reduction of phase noise is observed even for a relatively low driving force μ = $I_{rf}/I_{dc}$ = 0.02 (see orange curve in Fig. 2.b). In general, for such low injection power, we observe a plateau in the phase noise from a high frequency roll-off, $f_{roll-off}$, down to a low offset frequency corner $f_c$ (for example, for μ = 0.02 at $F_s = f_0$ in Fig. 2.b, $f_c$ = 200 kHz and $f_{roll-off}$ = 3 MHz). Similar behavior with different plateau levels is obtained for an external frequency at $2f_0$ (Fig. 2.c) or $f_0/2$ (Fig. 2.a) but for a larger external rf current, resp. μ = 0.08 and μ = 0.15.

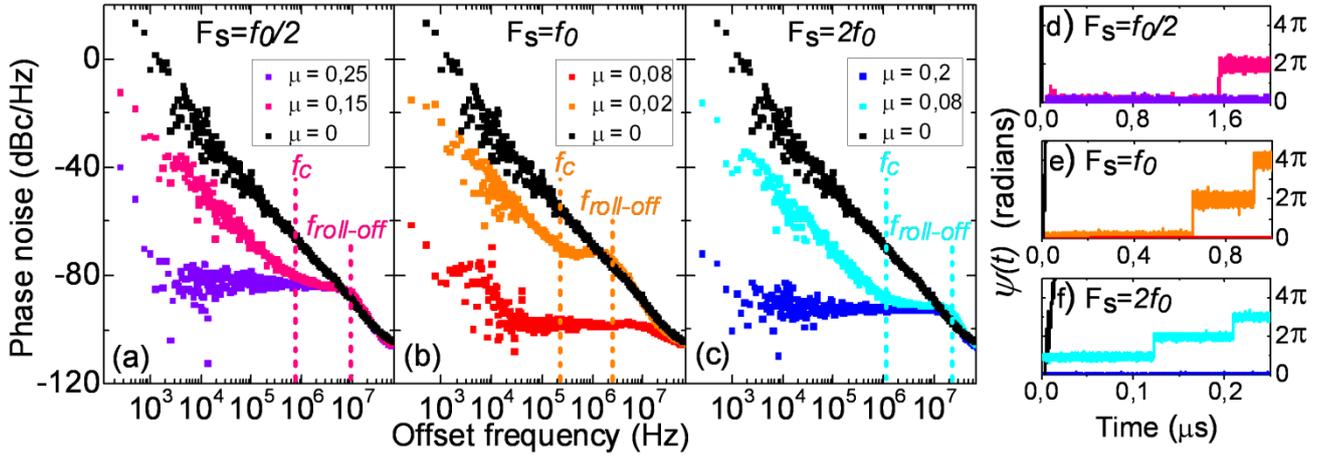

Fig. 2 : Phase noise of the locked oscillator for different driving force (at zero field, $I_{dc}$ = +11mA and Ø = 300 nm) at $f_0/2$ (a), $f_0$ (b) and $2f_0$ (b). Associated phase deviation ($\psi=\theta(t)-\omega_0 t$) for $F_s$ at $f_0/2$ (d), $f_0$ (e) and $2f_0$ (f).

For an offset frequency lower than $f_c$, the phase noise increases because of some phase slip events in the dynamics of our vortex based STNOs [15]. Indeed these features have been reported in micromagnetic simulations but never clearly observed in experiments [28]. These phase slips are associated to desynchronization-resynchronization events occurring due to the thermal energy. The amplitude of these phase slips is related to the number of stable positions over one period of vortex oscillations, which directly depends on the frequency of the current source. A direct consequence is that, for driving force at a $qf_0$ frequency, the phase deviation ($\psi(t) = \theta(t)-2\pi f_0 t$ with $\theta(t)$ the phase of the STNO) of the synchronized state presents phase slip events with an amplitude of $2\pi/q$ as demonstrated in Fig. 2.f for the $2f_0$ case (see also the $3f_0$ case in the supplementary information [22]). Thus for $F_s = f_0/2$ (or any other $f_0/q$), because the source frequency is lower than the one of the STNO, the resynchronization events present also a $2\pi$ phase slip as shown in Fig. 2.d. An important conclusion is thus to clearly differentiate a "frequency locking" state and a real phase locking state as these phase slip events are observed even when the oscillator frequency is locked to the external signal.

For large enough driving forces (in Fig. 2.a-c, resp. the μ = 0.08 red curve at $f_0$, the μ = 0.25 purple curve at $f_0/2$ and the μ = 0.2 dark blue curve at $2f_0$), a flat plateau of phase noise from the high $f_{roll-off}$ is observed down to a 300 Hz offset frequency (lowest value associated with our oscilloscope memory). In this "perfect" synchronized regime, the phase noise level is -90dBc/Hz at 1 kHz from the carrier when the external frequency is at $2f_0$ (see Fig. 2.c). Note that similar low phase noise is reached at $f_0$ (Fig. 2.b) but the data at very low offset frequencies are hindered by the intrinsic phase noise of our rf source. The presence of such a phase noise plateau arises from the presence of a retroaction force on the phase due to the locking forces [20]. The decay time associated to this retroaction process is characterized by the high frequency roll-off $f_{roll-off}$.

To understand the conditions of synchronization, we extend the general model of the auto-oscillator in the non-autonomous regime [17,26] to our case of interest, i.e., the vortex oscillator under an external rf current. In the phase-locked regime, we have to include all the different contributions of the spin transfer forces that are directly acting on the phase $\theta(t)$ of the STNO through an alternative current $J_{rf} \cos(\omega_s t)$. It is to be noticed that, in a simple double vortex based spin valve system [15,21], the circular symmetry gives rise to spin transfer forces that allows reaching a self-sustained regime but that are independent of the oscillator phase [29]. Thus phase locking to an external rf current was not achievable in such a system. Importantly, in the hybrid systems studied here, the additional spin torque components originating from the top uniform magnetic layer of the synthetic antiferromagnet (CoFe/Ru/CoFeB) play a crucial role. Indeed, because the magnetization of the SAF top layer does not have a circular symmetry, its associated spin transfer forces depend on the phase of the oscillator [30]. In Eq. 1, we give an expression of the two active components of spin torque due to these in-plane components of the spin polarization that are arising from the magnetization of the top SAF layer [30]:

$$\begin{cases} \overrightarrow{F_{Slon//}} = \Lambda_{SL//} J_{rf} \cos(\omega_s t) \begin{cases} \sin\theta(t)(\overrightarrow{u_\rho}) \\ \cos\theta(t)(\overrightarrow{u_\chi}) \end{cases} (1a) \\ \overrightarrow{F_{FL//}} = \Lambda_{FL//} J_{rf} \cos(\omega_s t) \begin{cases} -\cos\theta(t)(\overrightarrow{u_\rho}) \\ \sin\theta(t)(\overrightarrow{u_\chi}) \end{cases} (1b) \end{cases}$$

Where $\overrightarrow{u_\rho}$ and $\overrightarrow{u_\chi}$ are the polar vectors defined by the vortex core position (ρ,χ) [22] (with ρ the radius of oscillations and χ(t), the instantaneous angle of the vortex core position with the polarizer direction), $\Lambda_{SL//}$ and $\Lambda_{FL//}$ are the respective efficiency of Slonczewski and field like in plane torques. It should be noted that the oscillator phase $\theta(t)$ is defined by the oscillations of magneto-resistance, and so by the angle of the instantaneous in-plane magnetization of the vortex through the relation $\theta(t)=\chi(t)+C\pi/2$ (with $C=\pm 1$ depending on the vortex chirality [30]). The impact of those spin transfer forces on the auto-oscillating regime establishes a new and non-autonomous state for the oscillator. It results in the first order (in the regime of small perturbation) to power fluctuations $\delta p$ and a phase difference $\psi(t)=\theta(t)-\omega_s t$ between the source and the oscillator [17] :

$$(2) \begin{cases} \dfrac{d\delta p}{dt} = -2\Gamma_p \delta p + F\sqrt{p_0}\cos(\Psi + \Psi_{st}) \\ \dfrac{d\Psi}{dt} + \Delta\omega = +N\delta p - \dfrac{F}{\sqrt{p_0}}\sin(\Psi + \Psi_{st}) \end{cases}$$

with $\Gamma_p$ is the effective relaxation damping rate, $N$ the nonlinear frequency shift, $F = (\Lambda_{SL//}^2 + \Lambda_{FL//}^2)^{1/2} * J_{rf}/(2GR)$ the normalized external driving force, the additional locking force phase shift $\Psi_{st} = tan^{-1}(\Lambda_{FL//}/\Lambda_{SL//})$ for positive bias, $G$ the gyrotropic constant [30], R the dot radius, $\Delta\omega = \omega_s - 2\pi f_0$ the frequency detuning, $p_0$ the normalized power in the autonomous regime (that is directly related to the amplitude of gyration of the vortex core [26]).

Then we can determine the stable solution as:

$$\begin{cases} \delta p = p_0 \dfrac{\nu\Delta\omega + \sqrt{(1+\nu^2)F^2 - \Delta\omega^2}}{(1+\nu^2)\Gamma_p} & (3a) \\ \psi_0 = tan^{-1}(\nu) - sin^{-1}\left(\dfrac{\Delta\omega}{\sqrt{(1+\nu^2)}F}\right) - \Psi_{st} & (3b) \end{cases}$$

where $\nu=Np/\Gamma_p$ is the nonlinear dimensionless parameter [31]. These new equilibrium dynamics are valid inside the frequency locking bandwidth that depends on the strength of the locking force at the considered locking harmonic (for more details, see supplementary information [22]).

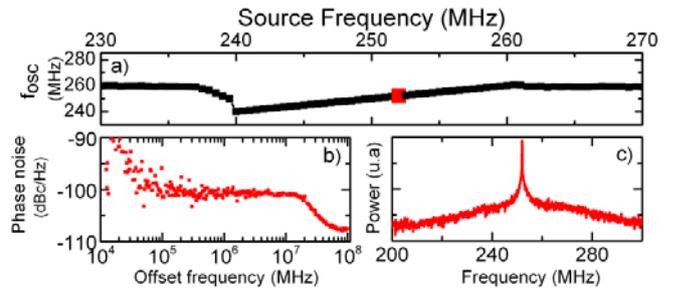

Fig. 4: Locking range at $f_0$ for μ = 0.2 (at zero field, $I_{dc}$ = +11 mA and Ø = 300 nm) (a) Phase noise spectrum (b) and its associated emitted spectrum (c) for $F_s$ = 252MHz

In the synchronized regime, an important parameter is the phase shift parameter $\psi_0$ (phase difference between source and oscillator phase) which should ideally be zero, if synchronizing multiple oscillators connected in series is targeted [13]. However given the generally large non-linearities of STNOs, i.e large ν, their dephasing $\psi_0$ is equal to $+\pi/2$ at zero detuning ($\Delta\omega = 0$) as reported in uniformly magnetized STNOs [20]. In vortex based STNOs, the implication of two forces in the locking process gives a unique opportunity to tune the phase shift $\psi_0$ through the additional term $\Psi_{st}$. Thus in magnetic tunnel junctions, $\psi_0$ could reach 0 at $\Delta\omega = 0$ for large ratio of $\Lambda_{FL//}/\Lambda_{SL//} = \xi R/b >> 1$ (with $\xi$ the efficiency of the field like torque, $R$ the dot radius and $b$ the vortex core radius [30]). This dephasing parameter $\psi_0$ also has some consequences in the transient synchronization regime through the decay rate of phase and power deviations of the stationary phase locked state [20]:

$$\lambda = \Gamma_p + \frac{1}{2}F\cos\psi_0 \pm \sqrt{\left(\Gamma_p - \frac{1}{2}F\cos\psi_0\right)^2 - 2\nu F \sin\psi_0} \quad (4)$$

The decay constant of the phase fluctuations can be related to the large frequency roll-off $f_{roll-off}$ displayed in Fig 2. Furthermore it should be noted that the decay constants of a synchronized nonlinear oscillator have a non-real part if $\psi_0$ is close to $+\pi/2$. Thus, in the presence of a complex decay, a synchronized state with sidebands is expected in the locking regime (as reported in uniformly magnetized spin-transfer nano-oscillators [19,20]). In Fig. 4.c, we clearly see that such sidebands are not observed for $F_s$ close to $f_0$. Moreover, on the phase noise diagram (Fig. 4.b), we note the absence of bumps or side bands at the high frequency roll-off $f_{roll-off}$. Both these features (correlated with Eq. 3.b and 4) are consistent with a large $\Lambda_{FL//}/\Lambda_{SL//}$ ratio in our vortex based MTJs so that a zero dephasing at $\varDelta\omega = 0$ can be expected depending on the sign of the field like torque [32].

In conclusion, we succeed to demonstrate the perfect injection locking of a vortex based spin transfer oscillator on an external current with multiple integer, i.e., $f_0$, $2f_0$ $3f_0$ as well as half integer frequency $f_0/2$. Ultra-low phase noise - 90dBc/Hz at 1 kHz offset and large output power (> 1 µW) in the locking state obtained at room temperature and zero field are reported. We demonstrate that the physical mechanisms at play for the oscillator to be locked, notably in the transient regime of synchronization, are strongly correlated to the symmetry of the spin transfer forces. The improved understanding of the locking behavior and the fine control of the oscillator phase allows envisaging the efficient synchronization of a large number of spin transfer oscillators, which would be a real breakthrough towards applications in advanced rf devices or novel family of neuro-inspired memories.

The authors acknowledge the ANR agency (SPINNOVA ANR-11-NANO-0016) as well as EU FP7 grant (MOSAIC No. ICT-FP7-8.317950) for financial support. E.G. acknowledges CNES and DGA for their support.

* now at INLN, Sophia Antipolis, Nice, France

+ now at IEF, CNRS, Orsay, France

## C. References

description of the amplitude of phase shift at $F_s=3f_0$ and of the impact of the locking strength on the synchronized equilibrium.

## Supplementary Materials:

### A) Phase slip for an rf current source at frequency $F_s$ close to $3f_0$, the oscillator frequency

As described in the main text, we show in Fig. 3.b that the level of phase noise strongly decreases in the locking regime compared to the auto-oscillating regime. Furthermore, we note the presence of phase slips for a locking strength µ=0.25, which corresponds to resynchronization events. Given that the source frequency is equal to three times the oscillator frequency, in one oscillator cycle there are three equivalent positions so that the amplitude of phase slip $2\pi/3$.

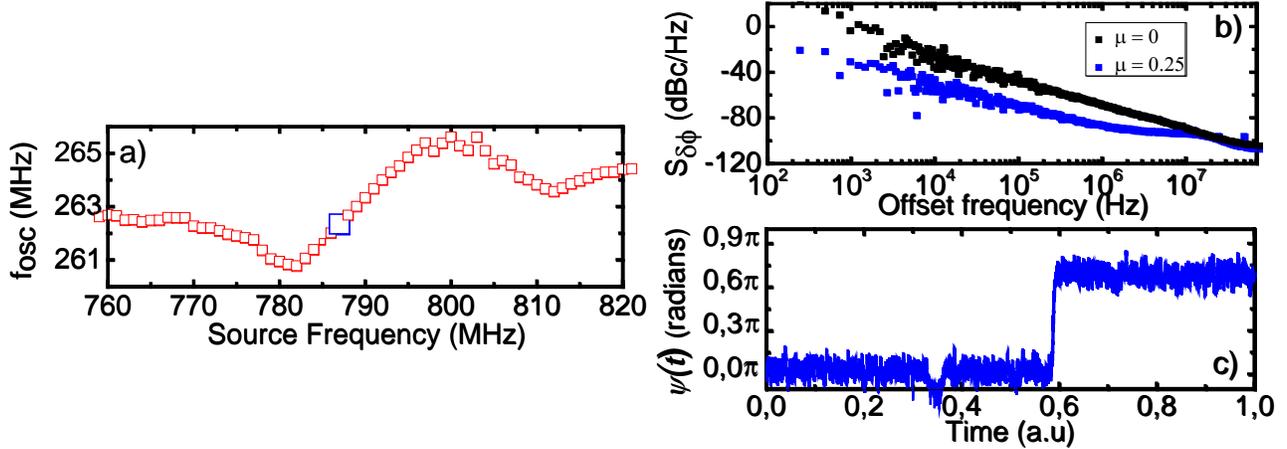

Fig. S1 (a) Synchronization at $F_s = 3f_0$ for $I_{rf} = 2.2$ mA and $I_{dc} = + 11$ mA and zero applied field (b) Phase noise at 787 MHz for $I_{rf} = 2.2$ mA and $I_{dc} = + 11$ mA (c) Associated phase deviation with a $2\pi/3$ phase slip

### B) Locking range characteristics depending on the locking force frequency and amplitude

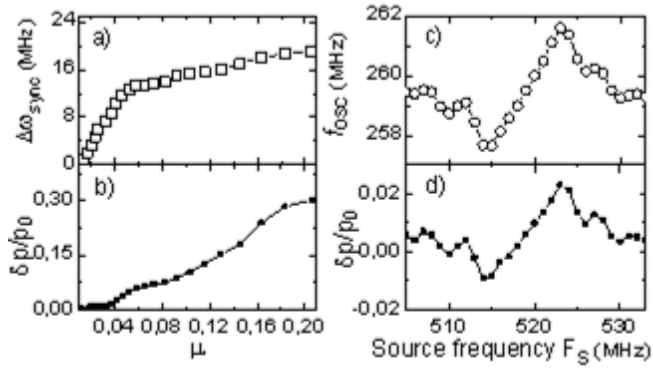

Fig. S2: Locking range at $2f_0$ (a) and normalized power $p_0$ for $F_s = 2f_0$ (b) dependence with the rf current for $I_{dc} = 11$mA. Oscillator frequency $f_{osc}$ (c) and normalized power $p_0$ (d) dependency with the source frequency $F_s$

The description of the phase locking regime is not limited to the case of an external current at the fundamental frequency $F_s = f_0$. It can be easily extended to any other ($f_0$:$qF_s$) frequency ratio through a parameter $g_{1,q}$ describing the efficiency of the driving force F at higher harmonic frequencies[1]. The (1:$q$) locking range can be thus defined with the classical expression [1], [2]

$$\Delta\omega_{sync} = \frac{\sqrt{(1+v^2)}F}{\sqrt{p_0}} g_{1,q} \quad (4)$$

In Fig. 3.a-b of the main text, we display how the experimental locking range (for an external frequency at $2f_0$) and the normalized power evolve with µ, the normalized external rf current. For µ < 0.05, we find a linear dependence of the locking range as predicted from Eq. 4 (of the main text) which then changes slope for a large driving force. As observed on Fig. 3.b, this deviation from the linear dependency expected from our model might be related to the significant increase, i.e. a non-small perturbation, of the normalized power $\delta p$ with μ (or $\delta p$ with $F$ in Eq. 3.a of the main text), which can also impact υ. Note that this dependence of the locking range with $p_0$ indicates that parametric synchronization can be neglected in our experiments[3]. Furthermore, as predicted from Eq. 3.a of the main text and shown in Fig. 3.c-d, the vortex oscillator adapts its frequency to the external frequency through a substantial modification $\delta p$ of its normalized power $p_0$ and hence its gyration radius. These two features illustrate the amplitude/phase coupling of our nonlinear oscillator.